\documentclass[fleqn,usenatbib]{mnras}

\usepackage{newtxtext,newtxmath}

\usepackage[T1]{fontenc}
\usepackage{ae,aecompl}

\usepackage{graphicx}	
\usepackage{amsmath}	
\usepackage{amssymb}	

\title[Turbulence profile limitations on adaptive optics tomography]{Limitations imposed by optical turbulence profile structure and evolution on tomographic reconstruction for the ELT}

\author[O. J. D. Farley et al.]{
O. J. D. Farley,$^{1}$\thanks{E-mail: o.j.d.farley@durham.ac.uk}
J. Osborn,$^{1}$
T. Morris,$^{1}$
T. Fusco,$^{2,3}$
B. Neichel,$^{2}$
C. Correia$^{2,4}$
\newauthor and R. W. Wilson$^{1}$
\\
$^{1}$Centre for Advanced Instrumentation (CfAI), Dept. Physics, Durham University, Durham, DH1 3LE, UK\\
$^{2}$Aix Marseille Univ, CNRS, CNES, LAM, Marseille, France\\
$^{3}$ONERA, 29 avenue de la Division Leclerc, 92322 Ch\^{a}tillon, France\\
$^{4}$W. M. Keck Observatory, 65-1120 Mamalahoa Highway, Kamuela, HI 96743\\
}

\date{Accepted XXX. Received YYY; in original form ZZZ}

\pubyear{2020}

\begin{document}
\label{firstpage}
\pagerange{\pageref{firstpage}--\pageref{lastpage}}
\maketitle

\begin{abstract}
The performance of tomographic adaptive optics systems is intrinsically linked to the vertical profile of optical turbulence. Firstly, a sufficient number of discrete turbulent layers must be reconstructed to model the true continuous turbulence profile. Secondly over the course of an observation, the profile as seen by the telescope changes and the tomographic reconstructor must be updated. These changes can be due to the unpredictable evolution of turbulent layers on meteorological timescales as short as minutes. Here we investigate the effect of changing atmospheric conditions on the quality of tomographic reconstruction by coupling fast analytical adaptive optics simulation to a large database of 10 691 high resolution turbulence profiles measured over two years by the Stereo-SCIDAR instrument at ESO Paranal, Chile. This work represents the first investigation of these effects with a large, statistically significant sample of turbulence profiles. The statistical nature of the study allows us to assess not only the degradation and variability in tomographic error with a set of system parameters (e.g. number of layers, temporal update period) but also the required parameters to meet some error threshold. In the most challenging conditions where the profile is rapidly changing, these parameters must be far more tightly constrained in order to meet this threshold. By providing estimates of these constraints for a wide range of system geometries as well as the impact of different temporal optimisation strategies we may assist the designers of tomographic AO for the ELT to dimension their systems. 


\end{abstract}




\section{Introduction}
Turbulence in the Earth's atmosphere poses a fundamental limitation on ground-based astronomical observations. Both the phase and amplitude of incoming starlight are aberrated by random refractive index fluctuations of the atmosphere along the line of sight, leading to the effect known as seeing. In adaptive optics (AO), the phase aberrations are measured by a wavefront sensor (WFS) and corrected by a deformable mirror (DM) on millisecond timescales, with the aim of reducing the effect of the atmosphere and achieving greatly improved photometry, astrometry and contrast in the case of coronography.

In recent years, techniques have been employed in AO using multiple WFSs to measure the phase aberrations along several lines of sight. These measurements are combined using a tomographic algorithm to reconstruct the phase aberration along any individual line of sight within the field of view, or across the entire field of view. This allows correction of the atmosphere using one or several DMs, forming a family of tomographic AO techniques \citep[see e.g.][]{Rigaut2002, Beckers1988, Vidal2010, Rigaut2018}.

These tomographic AO systems are fundamental to the operation of the next generation of extremely large telescopes (ELTs) and as such it is important for the science goals of these telescopes that they operate to the fullest extent of their capability: applying the best possible atmospheric correction in the given conditions. This responsibility lies with the ``soft" real-time control (SRTC) system or supervisor, which takes telemetry (WFS measurements) from the AO system and maintains an up-to-date tomographic reconstruction matrix. The two key pieces of information to compute this matrix or reconstructor  are the system geometry, i.e. relative positions of natural and laser guide stars (NGS/LGS), as well as the vertical optical turbulence profile. The profile, parameterised by the refractive index structure constant $C_n^2(h)$, describes the distribution of turbulence as a function of altitude $h$. Imperfect tomographic reconstruction leads to tomographic error, which is common to all tomographic systems and forms a significant part of the total error budget \citep{Gilles2008, Martin2017}. 

The computation performed by the SRTC usually consists of two stages. First, the turbulence profile $C_n^2(h)$ and other atmospheric parameters such as the outer scale $L_0(h)$ are obtained by fitting measured WFS measurements from the AO telemetry through a variant of the SLope Detection and Ranging (SLODAR) technique \citep{Wilson2002, Gilles2010, Cortes2012, Vidal2010}. The profile, which is comprised of $N$ discrete layers, must be of sufficient fidelity to model the true continuous $C_n^2(h)$ profile. Also, since the turbulence is a random statistical process, we must average WFS measurements over a time period $\delta t$ in order for the statistics to converge \citep{Martin2012}. During this time we must assume the profile is stationary.

Depending on the method the time taken to perform this fitting procedure can depend on the number of WFS measurements as well as the number of reconstructed layers $N$ \citep{Gratadour2018}, however this dependence can be alleviated by fitting a fixed number of reconstruction layers and using only the WFS measurements containing the most information about the profile \citep{Laidlaw2019}.

These parameters, including the discrete turbulence profile, are then used in the computation of the tomographic reconstruction matrix, which differs depending on the AO configuration. If there are altitude conjugated DMs as in multi-conjugate AO (MCAO), then the full turbulence volume must be explicitly reconstructed onto the $N$ reconstruction layers \citep{Fusco2001}. This computation time therefore depends on both the number of WFS measurements and $N$. If there are no altitude conjugate DMs, as is the case in laser tomographic AO (LTAO), ground layer AO (GLAO) and multi object AO (MOAO), the turbulent volume need only be implicitly reconstructed \citep{Vidal2010}, in which case the computation depends only on the number of WFS measurements, as well as system specific parameters such as the number of targets for MOAO.

The large number of WFS measurements at ELT scales (up to 90k WFS measurements in total) makes both of these stages computationally expensive, and as such efforts have been made to accelerate the SRTC using both GPU and CPU architectures \citep{Gratadour2018}, as well as applying novel numerical techniques to speed up the computation \citep{Ellerbroek2002, Doucet2018}. It is not only the raw computational power of the SRTC that may limit this computation time. Many of the matrices involved may be precomputed offline, in which case they are stored in a database which must be accessed and the data loaded before computation. While this reduces the computational overhead, these matrices can be up to multiple terabytes in size \citep{Arcidiacono2018} and this could therefore place constraints in other areas of the SRTC such as the memory.

Both the system geometry and the turbulence profile are temporally evolving and together they constrain the rate at which the SRTC must update the reconstructor $\Delta t$. Over the course of an observation, the geometry of the tomographic problem changes since natural guide stars (NGS) move in the field of fixed laser guide stars (LGS). Additionally, as the pointing angle of the telescope changes the heights of turbulent layers also change with airmass. The turbulence profile itself in the free atmosphere is essentially a meteorological phenomenon and it varies on similar timescales to weather: from long-term seasonal changes to unpredictable variation on timescales as small as several minutes.

For a particular observation, the effects of changing system geometry are mostly predictable and as such their impact on AO performance and therefore requirements for the SRTC can be estimated. This is not the case for the effect of the changing turbulence profile, where changes are random and unpredictable in nature. Assessment of the impact of sub-optimal reconstruction due to an evolving profile is more difficult since we must use a large number of turbulence profiles and assess the impact on the tomographic error statistically for a particular site. This is not feasible with conventional Monte Carlo AO simulation which requires long computation times for a single profile.

Previous work has studied the effect of sub-optimal tomographic reconstruction with limited sets of turbulence profiles. It has been shown using 11 high resolution turbulence profiles measured from balloon flights at Paranal that the number of layers required to maintain good performance is between $N=10$ and $N=20$, depending on the LGS asterism diameter \citep{Fusco2010}. It has also been shown that using more advanced compression methods can reduce this number further \citep{Saxenhuber2017}. Temporal effects have been investigated using a limited set of real turbulence profiles from La Palma by \cite{Gendron2014}. They showed that the reconstructor should be reoptimised on timescales of $\Delta t = 10$ minutes, as well as that the minimum averaging time should be at least $\delta t =5-10$ minutes.

Here, we employ a large database of 10 691 high-resolution, high-sensitivity optical turbulence profiles measured by the Stereo-SCIDAR instrument at ESO Paranal, Chile \citep{Shepherd2014, Osborn2018a}. By coupling these real turbulence profiles with fast analytical tomographic AO simulation, we assess the impact of sub-optimal reconstruction on the tomographic error in a statistical manner. This allows us to draw robust conclusions as to the number of required reconstruction layers $N$, the reconstructor update period $\Delta t$ and averaging time $\delta t$ for the Paranal observatory. By taking two contrasting nights we also investigate the effect of different temporal optimisation strategies. We dimension our simulations according ELT currently under construction atop nearby Cerro Armazones which will employ multiple tomographic AO systems \citep{Neichel2016, Roux2018, Morris2016}.


\section{Simulation} \label{sec:sim}

An analytical Fourier-domain AO simulation \citep{Neichel2008} is used, allowing simulation times for a single profile at ELT scales of several seconds on modest hardware. With some parallelisation we are able to reach simulation times of around 20 minutes for all 10 691 profiles on an AMD EPYC 64-core server. 

We simulate a laser tomographic AO (LTAO) configuration, with a single ground-conjugated DM and six LGS in a circular asterism. The reconstructor is optimised and performance measured on-axis such that there is no contribution from generalised fitting error. In addition we only integrate the power spectral density within the AO-correction radius such that there is no contribution from classical fitting error. In this way we measure only the tomographic error associated with the imperfect reconstruction of the turbulent volume. Included in this tomographic error is a constant contribution from the noise in the LGS WFS measurements. 

We set the six LGS in asterisms with diameters $\Theta$ of 1, 2, and 4 arcminutes to obtain tomographic error estimates applicable to a wide range of instruments, from narrow-field LTAO to wide field MCAO and MOAO. We summarise the fixed simulation parameters in Table \ref{tab:sim_params}.

\begin{table}
    \centering
    \begin{tabular}{l|r}
    Telescope Diameter & 39.3 m  \\
    Projected subaperture size & 0.5 m \\
    Projected DM pitch & 0.5 m \\
    \# LGS & 6 \\
    \# DM & 1 \\
    DM conjugation altitude & 0 m \\
    Tomographic reconstructor & MMSE \\
    Outer scale $L_0$ & 25 m \\
    LGS noise & 1 rad$^2$ \\
    Zenith angle & 0 deg
    \end{tabular}
    \caption{Fixed simulation parameters for all LGS asterisms.}
    \label{tab:sim_params}
\end{table}

In most cases, we will compare tomographic error in the optimal case, where the profile is known perfectly, to the case where the reconstructor is sub-optimal. We will use the quantity $E$, the difference in quadrature in tomographic error  $\sigma_\mathrm{tomo}$ with sub-optimal parameters compared to the optimal case, i.e.
\begin{equation}
\begin{aligned}
    E(N, t-t_\mathrm{opt}, \delta t) = & \big[ \sigma_\mathrm{tomo}^2(N,t-t_\mathrm{opt}, \delta t) \\ & -  \sigma_{\mathrm{tomo}}^2(N=100,t-t_\mathrm{opt}=0,\delta t=0) \big]^{1/2},
\end{aligned}
\label{eq:E}
\end{equation}
for $N$ the number of reconstructed layers, $t-t_\mathrm{opt}$ the time since the previous reconstructor optimisation and $\delta t$ the time over which the profile is averaged at each reconstruction step. Therefore a value of $E=0$ occurs when the profile is known perfectly, and any increase corresponds to the additional incurred tomographic error with a sub-optimal reconstructor. 


It should be noted that the true optimal reconstructor averaging time is not 0 but some finite time $\delta t_{\mathrm{min}}$ since turbulence profiles are themselves measured from a temporal average of measurements. For the Stereo-SCIDAR at Paranal, $\delta t_{\mathrm{min}} \approx 140$ s. For the purpose of our analysis we must assume that $\delta t_{\mathrm{min}}$ is close enough to 0 as far as concerns the tomographic error such that we may neglect this aspect.

It is not only the increase in tomographic error from the optimal case that is of interest. At the other end of the scale, a useful quantity is the worst case performance. We define this as the tomographic error obtained if we ignore the variability of the atmosphere, and choose instead to precompute our tomographic reconstructor with a single turbulence profile. We have several options when selecting this profile. The absolute worst performance is obtained if we have no a priori information about the profile at all, which corresponds to an optimisation profile of constant $C_n^2(h)$. Unfortunately, performance with this profile is so poor (median values of $E$ between 160 and 520 nm rms depending on the LGS asterism diameter) that it does not provide a useful benchmark with which to compare our results --- regardless of our reconstructor parameters we always perform better than this.

\begin{figure}
    \centering
    \includegraphics{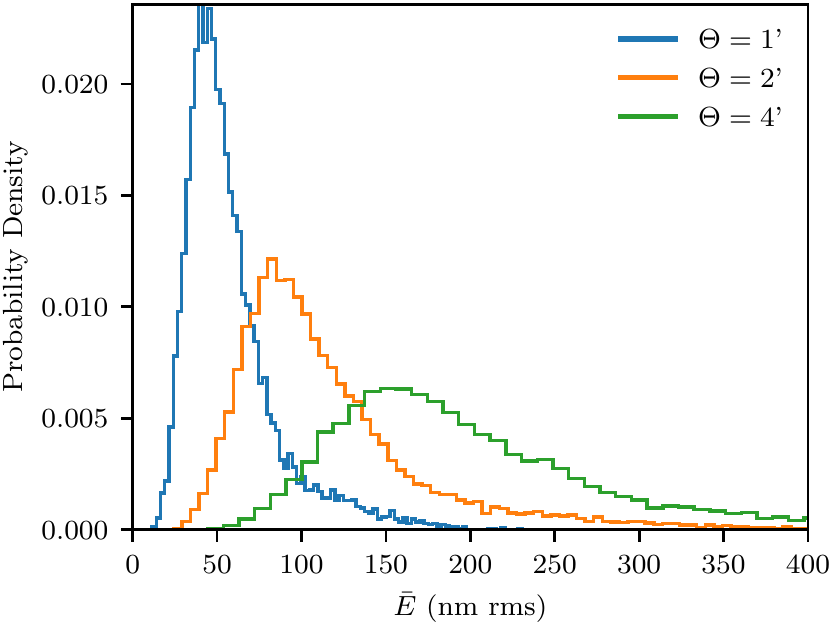}
    \caption{Distributions of $\bar{E}$ across the Stereo-SCIDAR 2018A data set obtained if we use only a single reconstructor computed with the ESO 35 layer profile, for 1, 2 and 4 arcminute LGS asterisms.}
    \label{fig:R_mean}
\end{figure}

A more reasonable prior for the turbulence profile is the 35 layer profile defined by ESO \citep{Sarazin2013}. This profile is designed to represent median conditions at Paranal and is often used in ELT simulations. In Fig. \ref{fig:R_mean}, we show the distributions of $E$ obtained if we optimise using this profile, which we denote $\bar{E}$. Quartiles of the distributions of $\bar{E}$ are also listed in Table \ref{tab:R_mean_table}. Similar distributions are obtained if we instead use a median  or mean profile over the 2018A dataset. We therefore conclude that these distributions represent approximately the best possible performance obtainable for a single profile reconstructor that is never updated.




\begin{table}
    \centering
    \begin{tabular}{c|ccccc}
        $\mathbf\Theta$ &$\mathbf P_5$ & $\mathbf P_{25}$ &$\mathbf P_{50}$ & $\mathbf P_{75}$ & $\mathbf P_{95}$ \\
        1 arcmin & 27 & 40 & 51 & 67 & 118 \\
        2 arcmin & 55 & 79 & 101 & 135 & 234 \\
        4 arcmin & 100 & 143 & 184 & 246 & 425
    \end{tabular} 
    \caption{Percentiles $P_i$ of the distributions of $\bar{E}$ shown in Fig. \ref{fig:R_mean}. Units are nm rms.}
    \label{tab:R_mean_table}
\end{table}

\begin{figure*}
    \centering
    \includegraphics{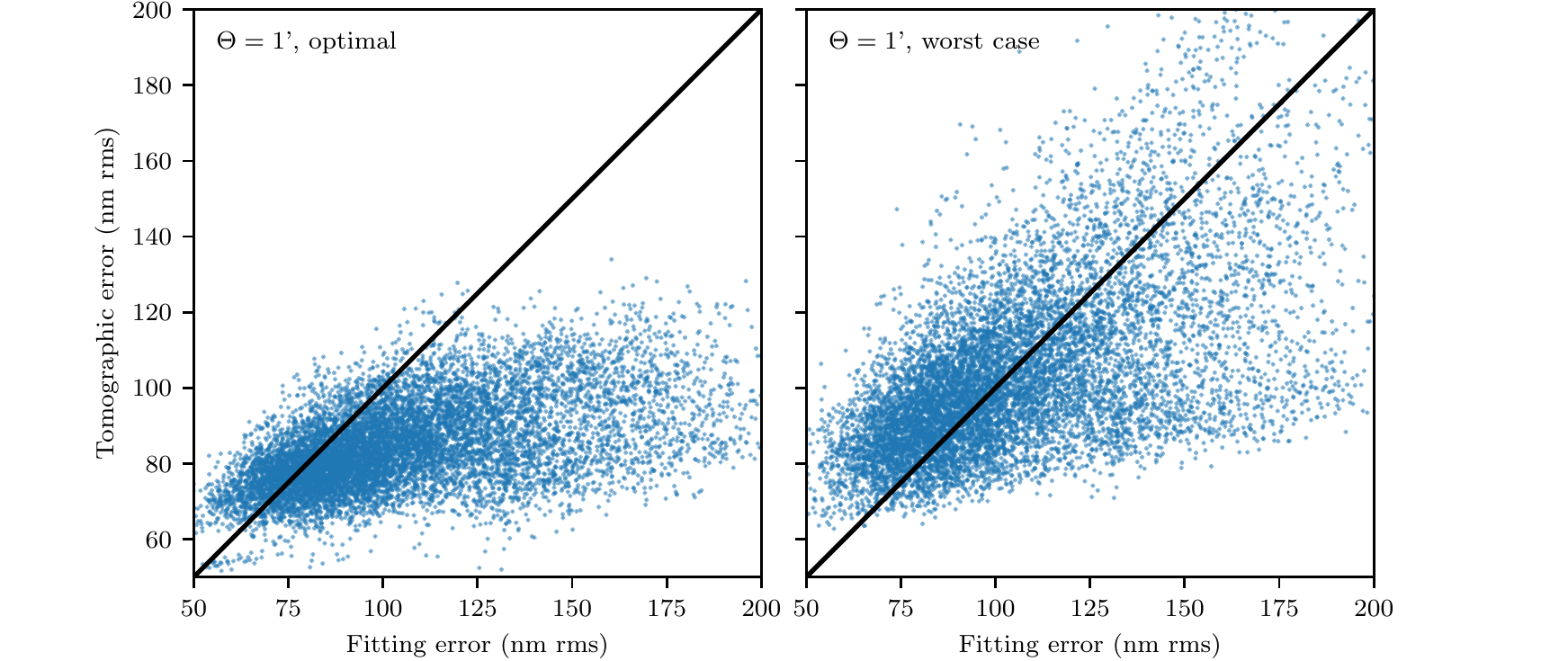}
    \caption{Tomographic error vs. fitting error for 1 arcminute LGS asterism, in the cases where the reconstructor is optimal (left) and the worst case (right). Each dot corresponds to one turbulence profile in the 2018A dataset. Black line corresponds to the case where the two errors are equal.}
    \label{fig:wfe_vs_fitting}
\end{figure*}

We focus on the tomographic error to maintain generality, however of course the overall performance of a tomographic AO system is defined by the sum of all error terms. Whether the performance impact of our worst case in Fig. \ref{fig:R_mean} is important will depend on the magnitude of these additional errors. Many of these errors will be highly system specific and as such would not be meaningful to include them here since we must make many assumptions as to the parameters of AO systems still in the design phase. However, in order to give some context to the values of tomographic error we can compute and compare to the DM fitting error over the 10 691 profiles. The fitting error can be computed as $\sigma_{\mathrm{fitting}}^2 = 0.23 (d/r_0)^{3/5}$ for $d$ the projected DM pitch and $r_0$ the Fried parameter \citep{Rigaut1998} and describes the inability of the DM to correct high order turbulent modes. We compute this using an assumed $d=0.5$ m for the ELT, and compare to the tomographic error for an LGS asterism diameter of 1 arcminute, which corresponds to the narrow field LTAO case where these two errors are most likely to be the largest contributors. We can see in Fig. \ref{fig:wfe_vs_fitting} that when the tomographic error is optimal, we are usually dominated by the fitting error. Only around 20\% of the time is the tomographic error larger than the fitting error. In the worst case this increases to over 50\%. By ensuring the reconstructor is optimised we can greatly improve the tomographic error when it is poor ($>100$ nm rms), ensuring that it is smaller than the fitting error in most cases. 

For systems with larger LGS asterisms the absolute gain from optimising the reconstructor will be greater as can be seen in Table \ref{tab:R_mean_table} and Fig. \ref{fig:R_mean}, however this may be cancelled out by larger additional error terms, for example generalised fitting in the MCAO case. As these systems are developed the values of $E$ computed throughout this work may be compared to to magnitude of these system specific errors to determine the extent to which the turbulence profile may limit performance.


\section{Number of reconstructed layers} \label{sec:nlayers}

In any tomographic AO system, the atmosphere is modelled as a number $N$ of discrete turbulent layers. Since in reality the turbulence profile is a continuous function $C_n^2(h)$, the accuracy of this model for a given profile depends on the number of layers as well as their altitude. If the number of layers is too few or their altitudes incorrect, we induce a model error which results in an increase in tomographic error.  The amount of model error induced by modelling the profile with $N$ layers will depend on the system geometry (e.g LGS positions) as well as the profile itself -- profiles with strong, discrete turbulent layers lend themselves to modelling with a smaller number of layers whereas more continuous distributions of turbulence require a greater number of layers to achieve the same model error. 

The input profiles from the Stereo-SCIDAR are of very high resolution, with $N=100$ altitude bins between the ground and 25 km. We can assume that these profiles provide a very good model of the continuous $C_n^2(h)$ profile and may therefore be used directly in our analytical simulation to provide our optimal case. 

The profiles obtainable by the SLODAR-type analysis of ELT WFS measurements will have a maximum resolution approximately equivalent to that of the Stereo-SCIDAR \citep{Vidal2010}. However for MCAO in particular, reconstructing such a large a number of layers poses an SRTC challenge. The computation of the reconstructor depends on the explicit projection of the turbulence measured by each WFS onto the $N$ reconstruction layers. This means that matrices which are already large (of dimension up to 90k $\times$ 90k) must be generated $N$ times and multiplied together. Alternatively if the geometry may be fixed, such as by separating NGS and LGS control \citep{Gilles2008a} and fixing the reconstruction layer altitudes, some of these large matrices may be precomputed and stored in a database. In this case the limiting factors may not only be the computing power but the speed at which the SRTC can retrieve and multiply these huge matrices \citep{Arcidiacono2018}.

For other forms of tomographic AO without altitude conjugate DMs, there is no explicit reconstruction of the $N$ layers hence the computation of the reconstructor itself does not depend on the number of layers. However, the turbulence profile must still have sufficient layers to avoid model error. Depending on the specific method used to fit the $C_n^2(h)$ profile from WFS measurements, the number of layers may still be a factor in determining the reconstructor update time for these systems \citep{Gratadour2018, Laidlaw2019}. There is also interest from the perspective of simulating these systems as to the minimum number of layers required to model the atmosphere.




Here we probe the effect on tomographic error of reconstructing $N\leq20$ turbulent layers, whilst maintaining the atmosphere at full $N=100$ layer resolution. We compare cases where the altitudes of the small number of layers are allowed to vary to the case where the altitudes are fixed. The high resolution profiles are compressed using three methods. Firstly, the equivalent layers method described in \cite{Fusco1999}. In this method, the turbulence profile is split into $N$ slabs, with the $C_n^2 (h)$ values for each layer being the integral of $C_n^2 (h) \, \mathrm{d}h$ in each slab. The height of each layer is then set to the mean effective height $\bar{h}=[\int{C_n^2 (h) h^{5/3}\, \mathrm {d}h } \,/\, \int{C_n^2(h)\, \mathrm{d}h} ]^{3/5}$ of the layers in each slab. This has the effect of reducing the number of layers whilst conserving the isoplanatic angle $\theta_0$, an important parameter in adaptive optics.

The second compression method we use here is the optimal grouping method described in \cite{Saxenhuber2017}. This method was shown in end-to-end AO simulation with a limited set of turbulence profiles to give better performance than many other compression methods including equivalent layers. This method is more computationally demanding, and involves the minimisation of the cost function 
\begin{equation}
    F = \sum_{l=1}^N \sum_{k \in G_l} C_n^2(h_k) | h_k - h_l|
\end{equation}
where $k$ runs over a particular grouping of turbulent layers $G_l$. 

Finally, the case where we fix the $N$ reconstruction altitudes. These fixed altitudes are decided by optimal grouping compression of the average 2018A profile. This is essentially a simple re-binning of the $C_n^2$ profiles.

\begin{figure}
    \centering
    \includegraphics{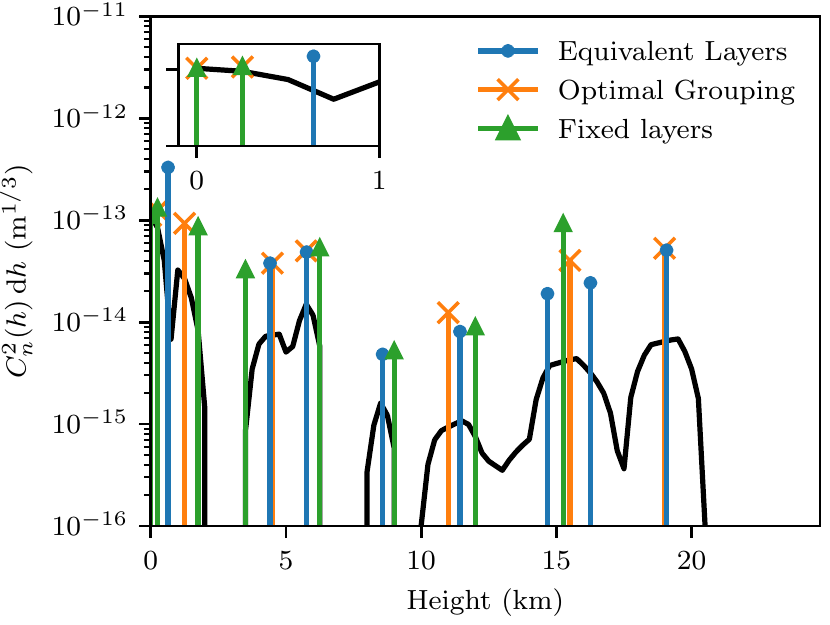}
    \caption{Use of the equivalent layers, optimal grouping and fixed layer methods to compress a high resolution Stereo-SCIDAR profile (black) measured on the night of the 8th May 2017. We compress from 100 layers to $N=8$ layers. Inset axis shows the detail in the ground layer ($h < 1$ km).}
    \label{fig:compression_egs}
\end{figure}

Application of these compression methods to an example profile can be seen in Fig. \ref{fig:compression_egs}. We can see that at some altitudes the layers from the methods overlap, whereas at other altitudes there is a disagreement between them. We perform this compression for all 10691 profiles in the 2018A data set for $N$ between 2 and 20 layers. 

\begin{figure*}
    \centering
    \includegraphics{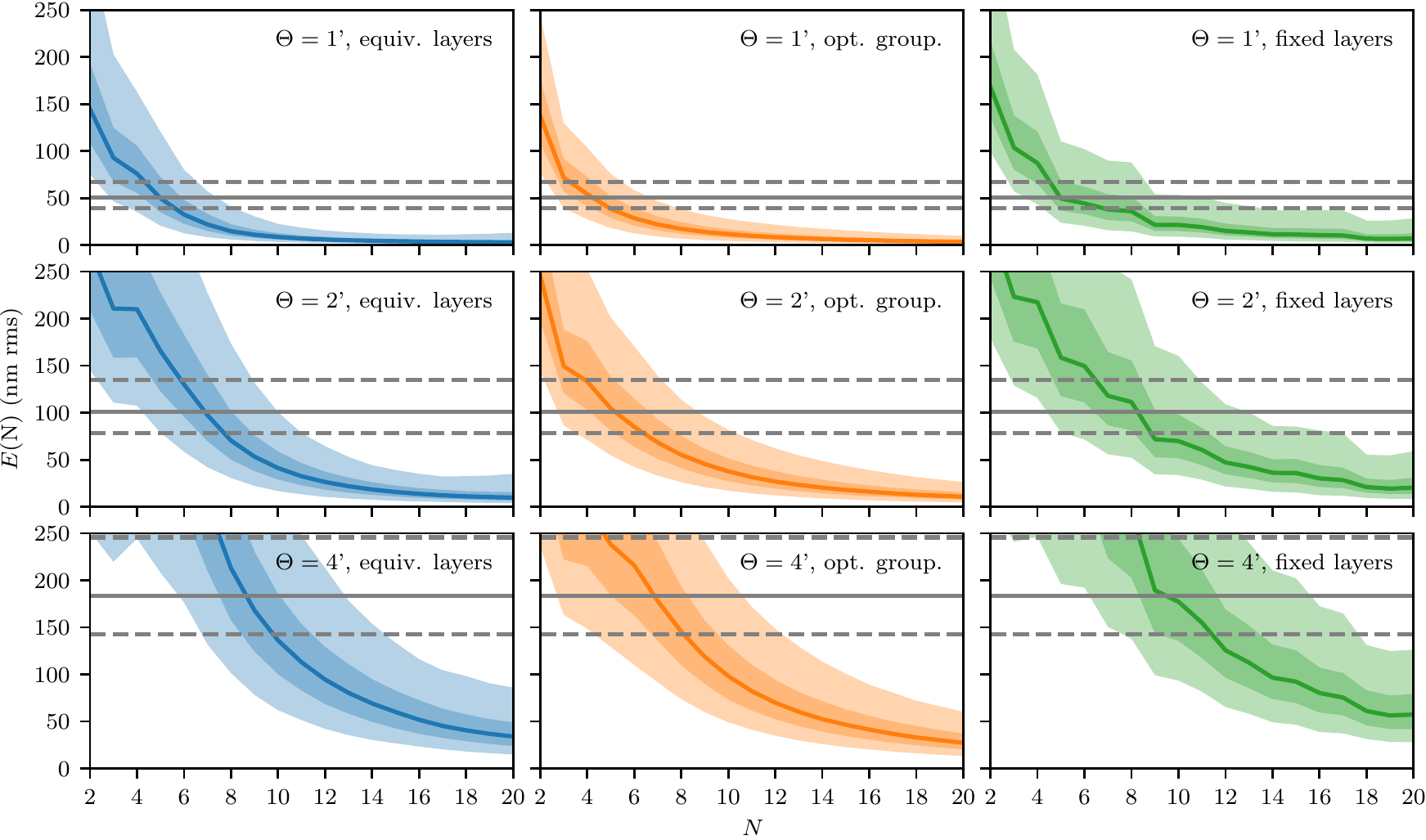}
    \caption{Tomographic error increase $E$ as a function of the number of reconstructed layers $N$. From upper to lower panels: 1, 2 and 4 arcminute diameter LGS asterisms. From left to right panels: equivalent layers, optimal grouping and fixed layer profile compression (blue, orange and green). In all panels the solid line indicates the median of the distribution over the 10 691 profiles. Darker shaded region indicates the interquartile range and lighter shaded region the 5th - 95th percentile range. Horizontal lines represent median, lower and upper quartiles of the worst case distribution of $\bar{E}$.}
    \label{fig:nlayers}
\end{figure*}

\begin{figure*}
    \centering
    \includegraphics{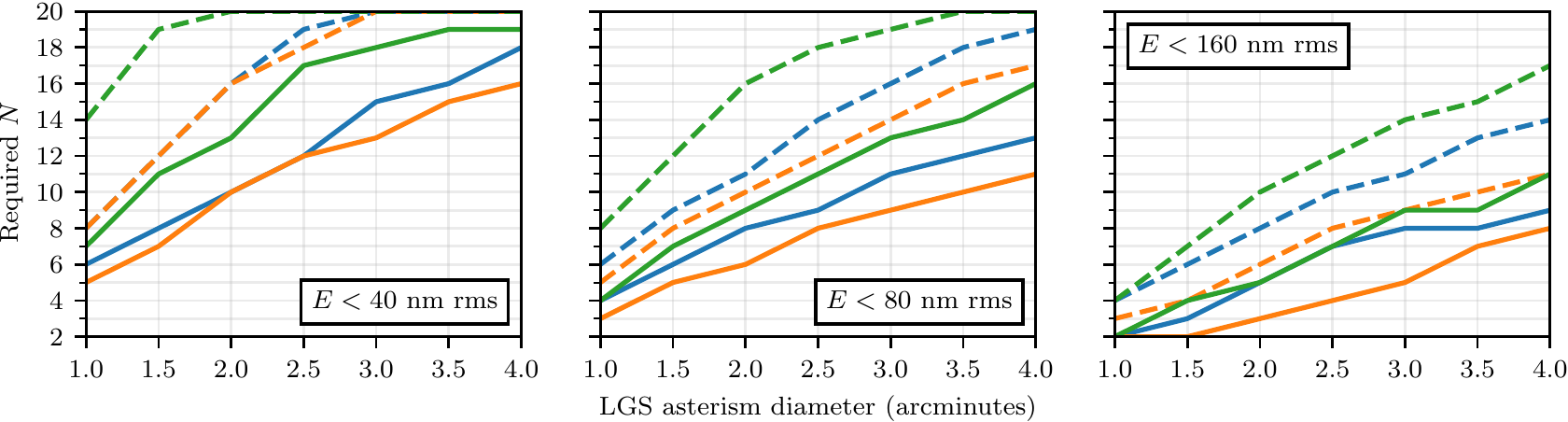}
    \caption{Number of layers required to reach maintain the increase in tomographic error below a tolerated threshold, with increasing LGS asterism diameter. From left to right: increasing error thresholds, indicated in the top-left corner. Blue represents profiles compressed with the equivalent layers method, orange are profiles compressed with the optimal grouping method and green are fixed layers. Solid line indicates the median and the dashed line the 95th percentile of the distributions over the 10 691 profiles for each asterism diameter.}
    \label{fig:nlayers_required}
\end{figure*}

In Fig. \ref{fig:nlayers} we illustrate how the tomographic error increases as the number of layers reconstructed is reduced for our 1, 2 and 4 arcminute LGS asterisms. We show here the spread of the distributions of $E(N)$ that result from the variation of the turbulence profile over the 10 691 measurements in the 2018A data set. It is immediately obvious that wider asterisms require a greater number of layers to be reconstructed to reach the same level of error. In addition the spread of the distribution increases with larger asterisms. The optimal grouping and equivalent layers methods show similar increases in tomographic error, with a slight advantage for the optimal grouping method for small $N$. In the 1 arcminute asterism case, we can reconstruct as few as 5 equivalent layers or 4 optimal grouping layers before the tomographic error reaches the worst case distribution $\bar{E}$. For fixed layers, the increase is larger as would be expected since the heights of the reconstruction layers are no longer matched to the current profile. This is most apparent in the 95th percentile of $E$ which is higher than for equivalent layers. This shows that there are a small set of profiles representing around 5\% of the time where using these fixed layers gives a particularly bad fit. 




The exact number of layers required depends on the error increase that can be tolerated for a particular system. In Fig. \ref{fig:nlayers_required} we show the distributions of the number of layers required to meet different error thresholds. Here we have additionally computed the number of layers required for some intermediate LGS asterisms as well as for 1, 2 and 4 arcminutes to better show the behaviour with increasing asterism diameter. If we tolerate a higher level of error, we are able to reconstruct fewer layers, and the optimal grouping method consistently allows between 1 and 4 fewer layers to be reconstructed depending on asterism diameter. The variability of turbulence profile can result in a large variability in the number of layers required. At larger asterism diameters there can be as much as a 5 layer difference between the median number of layers required and the 95th percentile.


\section{Temporal reconstructor optimisation}\label{sec:temporal}

\subsection{Tomographic error degradation over time} \label{sec:updaterate}
The tomographic reconstructor is computed using knowledge of both the system geometry and turbulence profile. If either of these change, the reconstructor will no longer be optimal and tomographic error will increase. 

Over the course of an observation, usually around one hour, several changes may occur in the geometry of the system as the telescope tracks the target across the sky. NGS, which must be used to recover wavefront tip-tilt information, may move with respect to the LGS due to field rotation. We do not consider errors of this type here since they are highly system specific.

\begin{figure}
    \centering
    \includegraphics{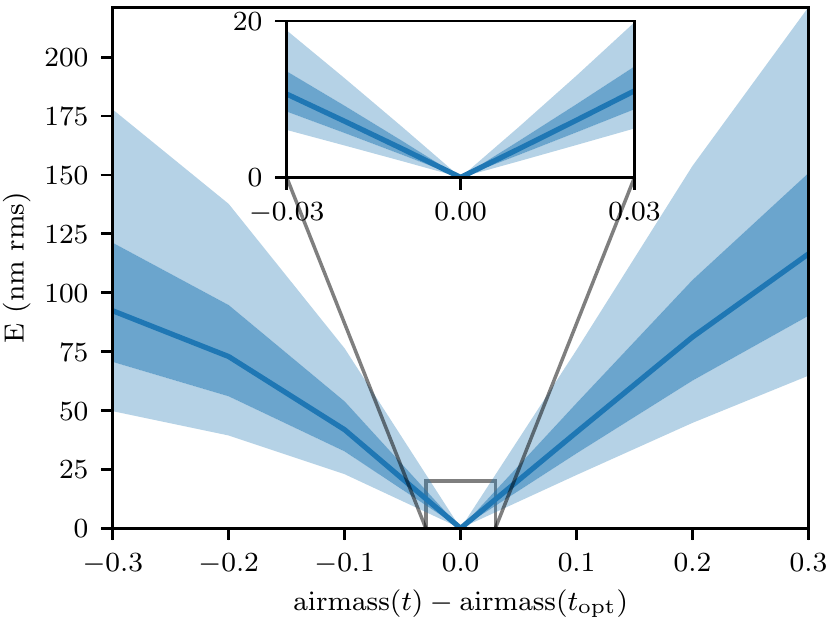}
    \caption{Increase in tomographic error with change in airmass over the course of a 1 hour observation if the reconstructor is not updated. We have simulated here only the 1 arcminute LGS asterism. Solid line indicates the median, darker shaded region the interquartile range and lighter shaded region the 5th - 95th percentile range over the 10691 profiles. The reconstructor is optimal at the origin ($t=t_\mathrm{opt}$), and we show the cases of both rising (negative change in airmass) and setting (positive change in airmass) targets. Primary axes show the worst observing far from zenith where the airmass changes considerably ($\pm 0.3$) over the observation. The inset axis shows a best case where the observation is near zenith and the change in airmass is 100 times smaller ($\pm 0.03$). }
    \label{fig:airmass}
\end{figure}

As the telescope zenith angle $\gamma$ changes the angle at which the turbulence profile is viewed changes, with the effective altitudes of each turbulent layer changing according to airmass $\sec{\gamma}$. The level of this error over an observation will depend on the target altitude, however we show best and worst cases over an approximately 1 hour observation in Fig. \ref{fig:airmass}. We can see that when observing near zenith, the error is very small and always less than 20 nm rms in the 1 armcinute asterism case. This is contrasted by the case where we observe far from zenith and as a result the change in airmass is large. Here we see large increases in tomographic error of the order of 100 nm after 1 hour. This error can be removed for an LTAO system by modifying the LGS asterism diameter in order to maintain constant geometry \citep{Neichel2016}. The specific implications for the tomographic error will depend on the telescope observation patterns and is beyond the scope of this work. 

We consider here the case where the telescope will be observing near zenith and as such the increase in error over time from changing airmass is small. What remains is the effect of the temporal evolution of the turbulence profile itself. Unlike the temporal errors above, this is not predictable over the course of an observation since the evolution of the profile is a meteorological phenomenon.

\begin{figure}
    \centering
    \includegraphics{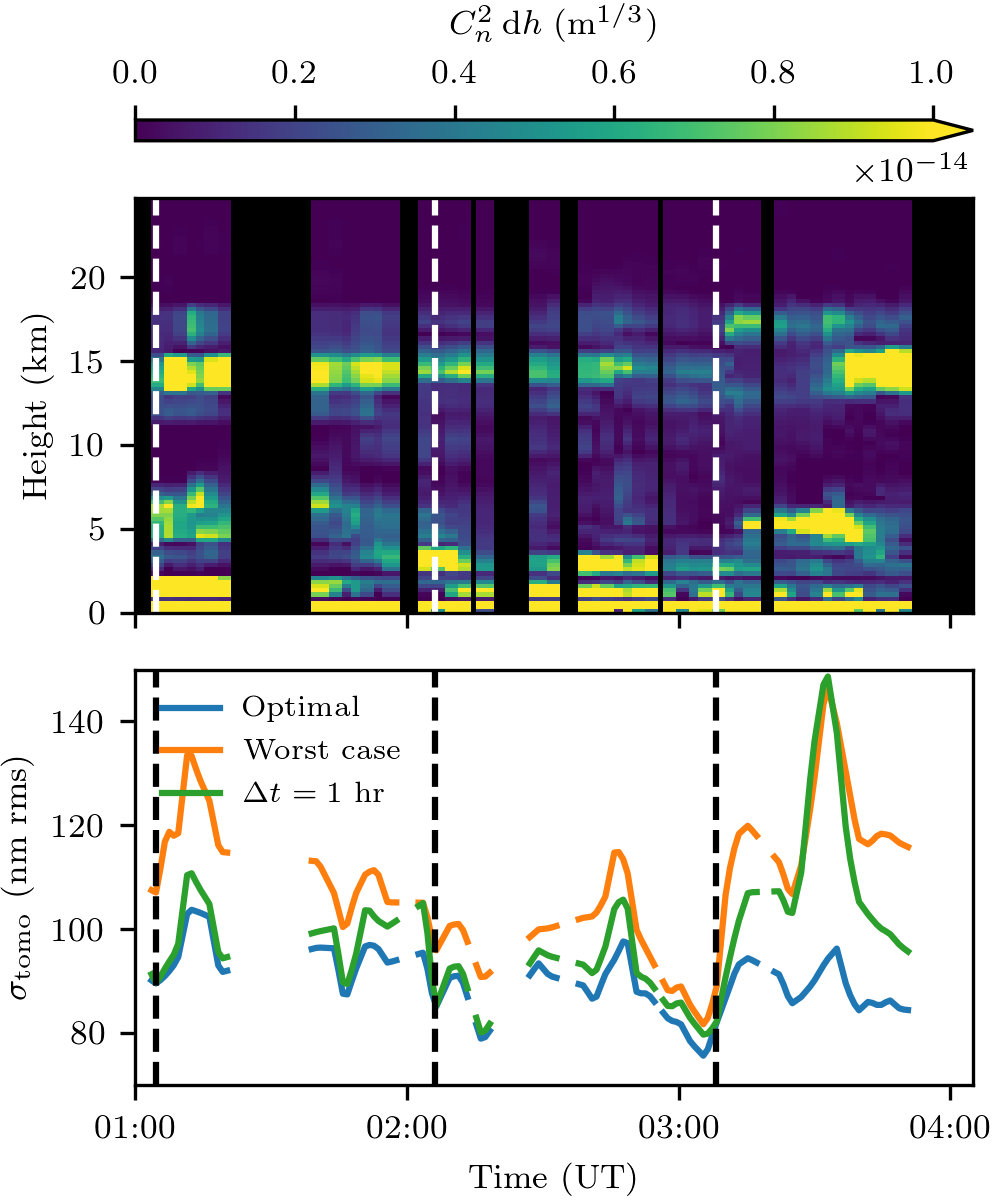}
    \caption{\textit{Upper}: Sequence of approximately 3 hours of Stereo-SCIDAR profiling data from the night of the 8th July 2017. The $C_n^2 \, \mathrm{d}h$ colour scale is clipped at $10^{-14}$ to emphasise high altitude layers. \textit{Lower}: Corresponding tomographic errors for the 1 arcminute asterism. In blue we show the optimal case where the profile is always perfectly known. In orange, the case where we do not optimise the reconstructor, and optimise only using the ESO 35 layer profile. Finally in green the tomographic error in the case when we update the reconstructor once per hour. Reconstructor optimisation times are indicated by vertical dashed lines.}
    \label{fig:wfe_jump}
\end{figure}

In some cases, as illustrated in Fig. \ref{fig:wfe_jump}, the profile may change dramatically from the profile that the reconstructor was last optimised with, leading to a large increase in tomographic error. In this particular example, at 03:00 UT the optimisation profile contains some weak high altitude layers above 10 km. After only a few minutes, a very strong layer appears at 5 km, resulting in an increase in tomographic error from around 90 nm to 110 nm. This is followed at approximately 03:30 UT by a strong layer at 15 km, pushing up the tomographic error to 150 nm rms. While this is an extreme example, the longer time period between reconstructor optimisations the more likely it is that the atmosphere will change and performance will suffer.

\begin{figure}
    \centering
    \includegraphics{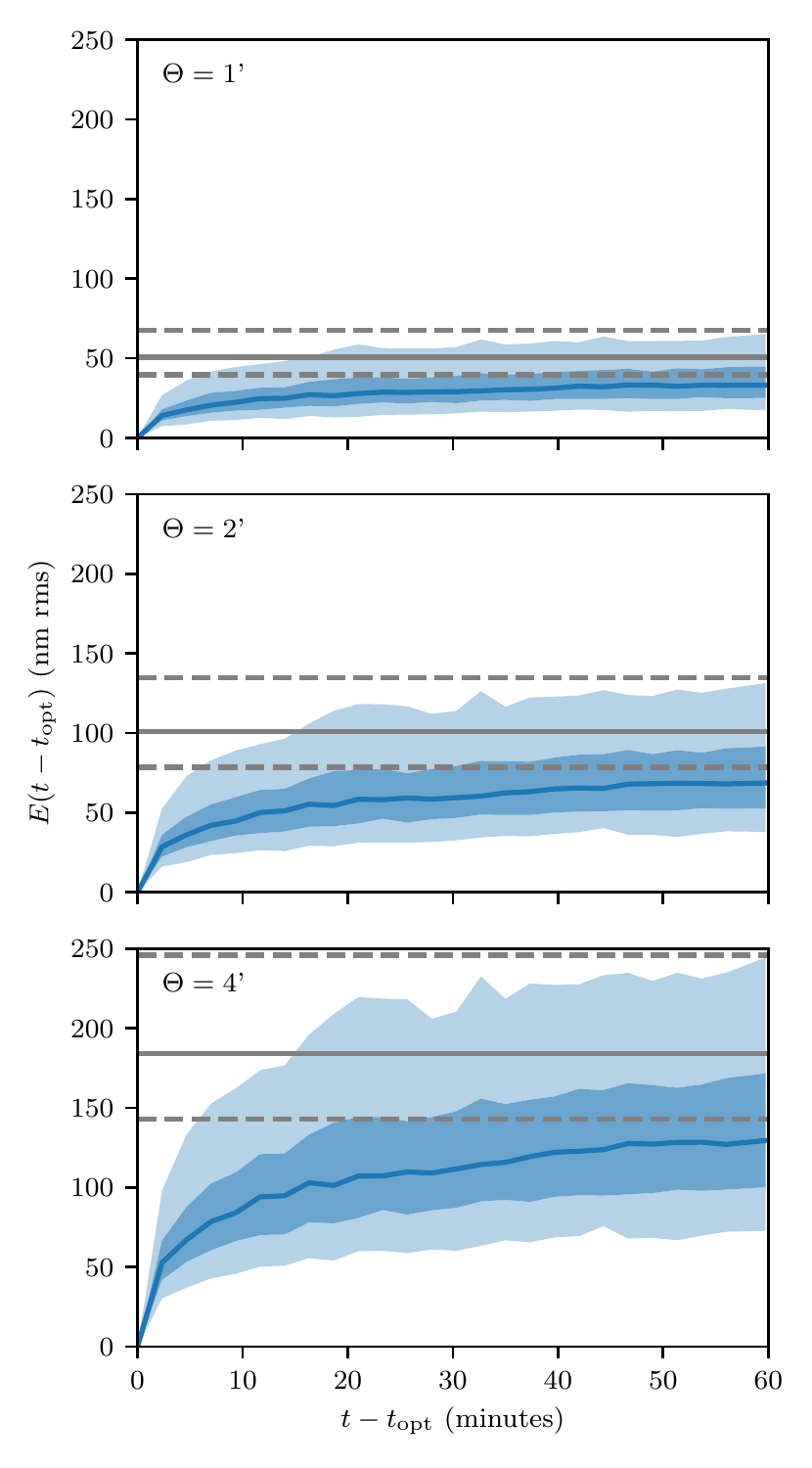}
    \caption{Tomographic error increase with increasing time since tomographic reconstuctor optimisation $t - t_{\mathrm{opt}}$. From upper to lower panel: 1, 2 and 4 arcminute LGS asterisms. The solid line indicates the  median  of  the  distribution  over  the  10  691  profiles.  Darker shaded region indicates the interquartile range and lighter shaded region the 5th - 95th percentile range.}
    \label{fig:temporal}
\end{figure}

To investigate this more generally, we update the reconstructor once per hour over all 10 691 profiles in the dataset and may ascertain the statistical behaviour of $E$ with increasing time since optimisation $t-t_\mathrm{opt}$. We show these results in Fig. \ref{fig:temporal}.

\begin{figure}
    \centering
    \includegraphics{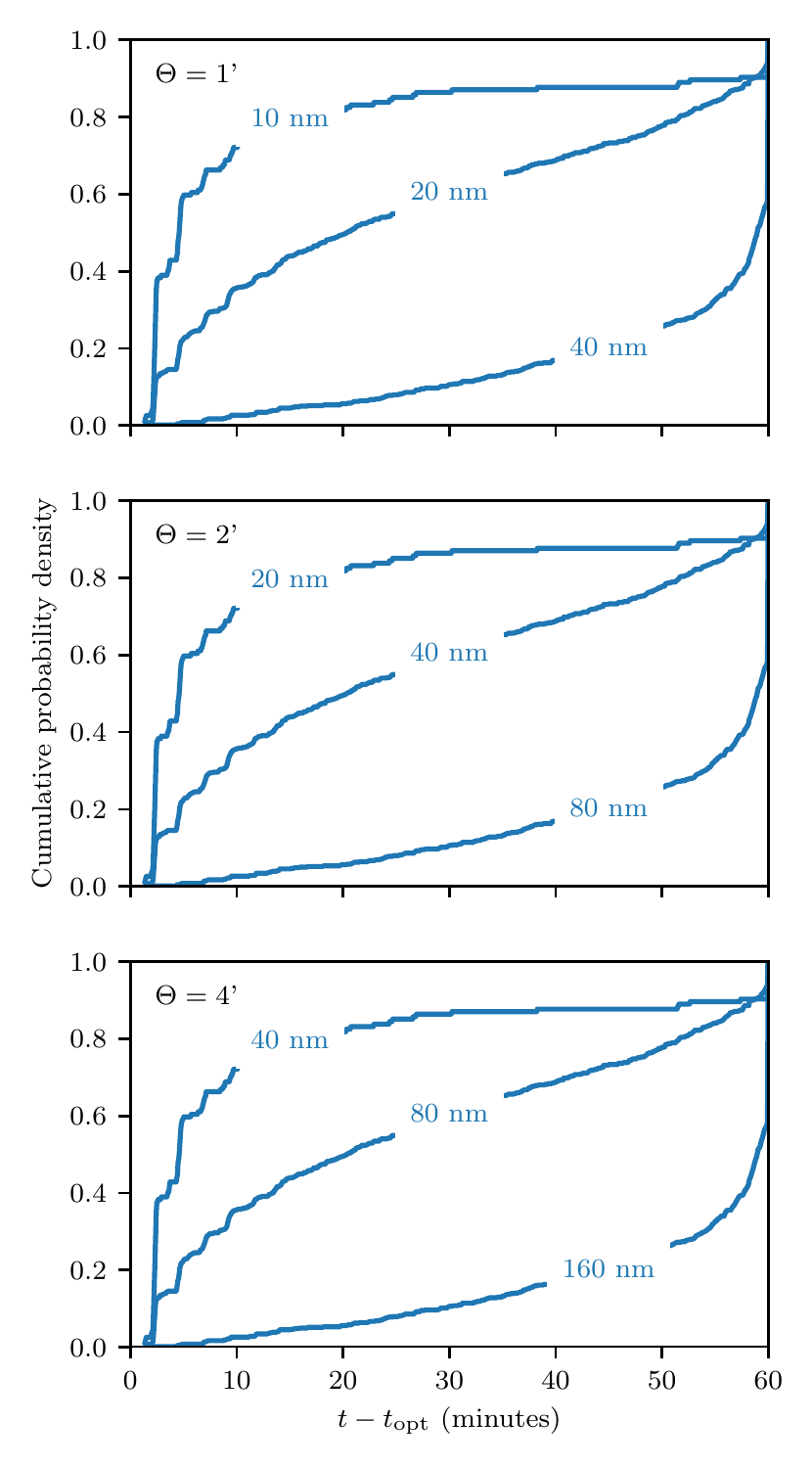}
    \caption{Cumulative probability distributions describing the time since the last optimisation step $t-t_\mathrm{opt}$ at which the increase in tomographic error reaches some threshold. From upper to lower panel: 1, 2 and 4 arcminute LGS asterisms. Each line is labelled with its $E_\mathrm{crit}$ threshold.}
    \label{fig:lessthan_Ecrit}
\end{figure}

The shape of the distributions over time are very similar for each LGS asterism, with larger absolute values of $E$ for increasing asterism diameter. Most of the additional tomographic error occurs on average in the first 10 to 20 minutes after optimisation of the reconstructor. After this point, the percentiles of the distribution plateau and there is a smaller increase for the remaining 40 minutes before the next optimisation step. After 60 minutes, the distributions do not reach the level of $\bar{E}$ as shown by the horizontal lines.

With the same data, we may also calculate another interesting set of distributions. If we wish to limit the tomographic error to some fixed increase $E<E_\mathrm{crit}$, we can calculate for each optimisation period the time at which we reach this threshold. In Fig. \ref{fig:lessthan_Ecrit}, we show that this time depends on the chosen error threshold and the LGS asterism diameter. As one would expect from the wide distributions in Fig. \ref{fig:temporal}, the time at which we meet this error threshold can vary dramatically. Taking for example the 1 arcminute asterism, we find that in the median case we reach an error threshold of $E_\mathrm{crit}=20$ nm rms in approximately 20 minutes. However across the data set this time ranges from less than 5 minutes (~10\% of profiles) to over 1 hour (~5\% of profiles). This makes it difficult to select a single optimisation period $\Delta t$.

\subsection{Reconstructor averaging time} \label{sec:avgtime}

Fitting the turbulence profile to WFS measurements requires average of many measurements must be used to minimise statistical noise and ensure the convergence of covariance matrices \citep{Martin2012}. During this averaging time the profile is assumed to be stationary however we know that sudden changes to the profile can occur on minute timescales. The profile measured  by the system is therefore the mean over the averaging time \citep{Gendron2014}. It has been shown that averaging profiles on a large scale (i.e. entire datasets) produces unrealistic continuous distributions of turbulence that provide poorer tomographic error than expected \citep{Farley2019}.


This averaging effect will result in an immediate degradation in performance, since the profile is no longer matched to the current profile at time $t=t_\mathrm{opt}$. However, an additional consequence of the averaging is more stable performance over time: if for example the profile behaves strangely during the few minutes when the reconstructor is being optimised this can lead to a sudden performance hit when the profile returns to normal.

There is therefore a trade-off between absolute performance, which will deteriorate as we average over greater timescales, and more constant performance over time which could result in better performance over longer timescales. The overall effect on the tomographic error will depend on how much the profile changes over the course of an update period, so it is clear that $\delta t$ will be linked to $\Delta t$. Note that we are only taking into account changes in the profile here, and we do not take into account other effects that temporal averaging has on the reconstructor such as better estimation of WFS covariance matrices, leading to a more accurate reconstructor. In reality, these effects would more likely determine the averaging time $\delta t$.

We perform similar analysis to Section \ref{sec:updaterate}, updating the reconstructor once per hour over the 10 691 profiles. At each optimisation step, instead of using the most recent profile, we optimise on the average of profiles from the time $t_\mathrm{opt} - t \leq \delta t$ with values of $\delta t$ ranging from 10 minutes to 60 minutes in 10 minute steps. There are cases where there are very few measured profiles in the timeframe $t_\mathrm{opt}-t \leq \delta t$, for example at the beginning of each night. To avoid biasing the results the data from these particular cases is removed from the subsequent analysis. In the worst case this filtering removes approximately 40\% of the data, leaving 6545 profiles which is still a large enough sample for our statistical analysis.

\begin{figure}
    \centering
    \includegraphics{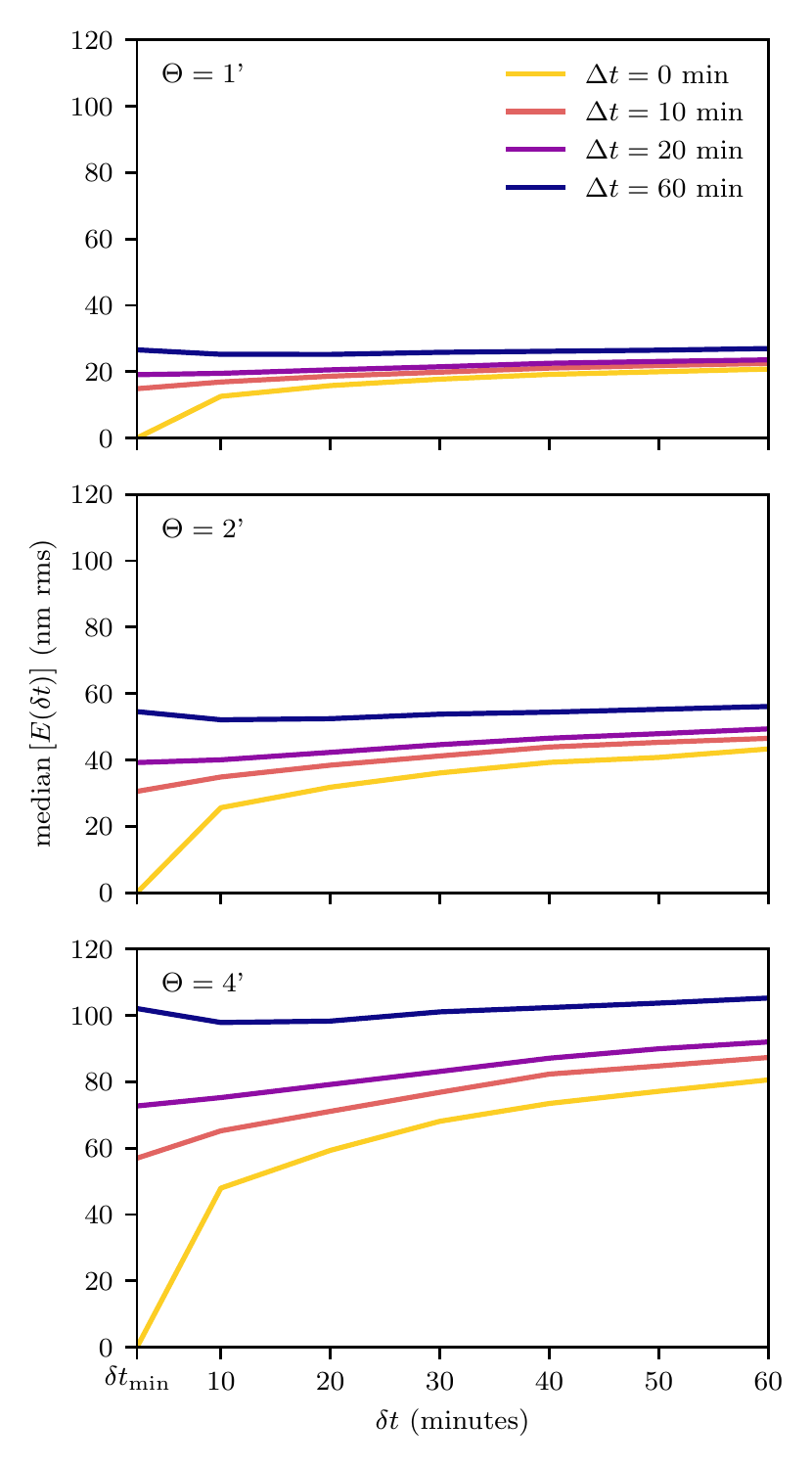}
    \caption{Median increase of tomographic error $E$ with averaging time $\delta t$. From upper to lower panel: 1, 2 and 4 arcminute LGS asterisms. Several different values of the reconstructor optimisation period $\Delta t$ are shown with different colours, from 0 to 1 hour. Horizontal grey lines indicate median (solid) and interquartile range (dashed) of worst case distribution $\bar{E}$.}
    \label{fig:delta_t}
\end{figure}

We present the results of this analysis in Fig. \ref{fig:delta_t}. For clarity we show only the median behaviour of $E(\delta t)$, but the distribution across the data set follows similar patterns. This behaviour is quite clearly dependent on the reconstructor update period $\Delta t$: when we are more frequently updating the reconstructor ($\Delta t < 20$ minutes), increasing the averaging time is detrimental. We are dominated in this regime by the increase in tomographic error resulting from the fact that an optimisation profile consisting of the mean over $\delta t$ is no longer matched to the current profile. If we increase $\Delta t$ and update the reconstructor less frequently, overall the performance deteriorates as would be expected from Fig. \ref{fig:temporal}. However the behaviour with $\delta t$ changes and there is some optimum averaging time $\delta t > \delta t_\mathrm{min}$ that provides slightly better overall performance. In this regime, for short averaging times we are dominated by increased error at large time since optimisation $t-t_\mathrm{opt}$. By increasing the averaging time we reduce this increase in error over time, resulting in very slightly better error overall. This averaging time is between 10 and 20 minutes for all LGS asterisms. Overall this effect is small and will only change the tomographic error by a few nm rms.

\section{Optimisation strategies} \label{sec:strats}

We have shown in section \ref{sec:updaterate} that unpredictable variability in the profile over time means it is difficult to select a single reoptimisation period for a particular asterism. Longer optimisation periods are a gamble with the atmosphere. If we are lucky and optimise at the correct time the profile will change very little and we will achieve near optimal performance. The opposite is also true, if we optimise at the wrong time then the tomographic error can rapidly degrade as seen in Fig. \ref{fig:wfe_jump}. Minimisation of these error spikes is important for a system to maintain consistent performance over an observation.


The risk of a sudden increase in error can of course be minimised by selecting the shortest possible reoptimisation period. We can see from Fig. \ref{fig:lessthan_Ecrit} that this should be less than around $\Delta t=10$ minutes (depending on the LGS asterism and tolerated error increase) to maintain near optimal performance in approximately 95\% of conditions. However, there is no reason why the optimisation period must be fixed. Indeed the unpredictable changes in the profile mean that there may be long periods of almost constant profile followed by rapid changes.  

Therefore we propose an alternative optimisation strategy, where the reconstructor is optimised not on some constant timescale $\Delta t$ but instead in the case where the increase in error $E$ reaches some threshold $E_\mathrm{crit}$. This can be accomplished in reality by employing the same analytical Fourier simulation used here. For a new turbulence profile measured by the system or an external profiler, the simulation can provide (in several seconds) an idea of how this new profile has affected the tomographic error of the system. If the error has degraded beyond $E_\mathrm{crit}$, then we optimise the reconstructor using this new profile.

\begin{figure*}
    \centering
    \includegraphics{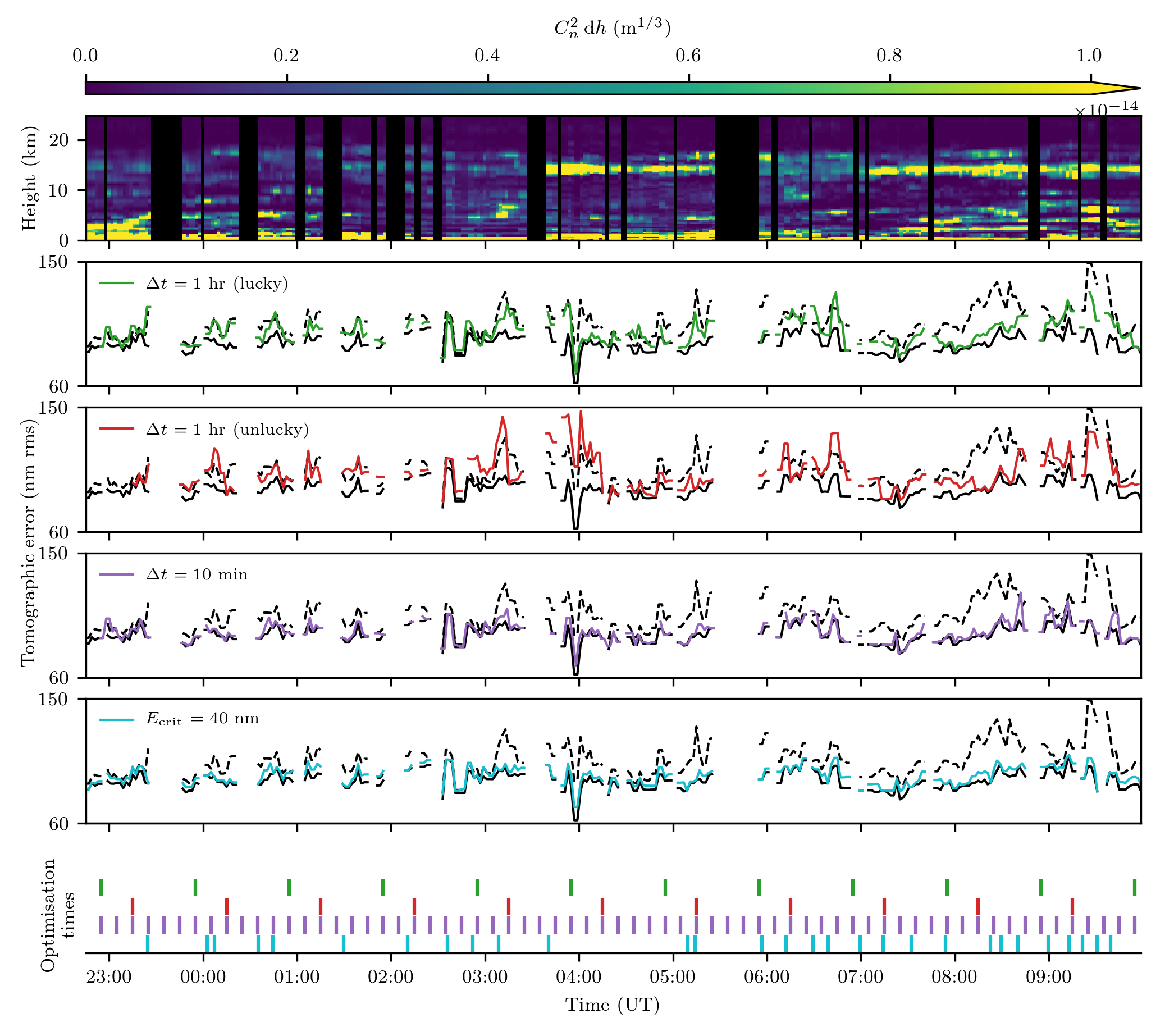}
    \caption{ELT tomographic error over the night of 6th August 2017, using a 1 arcminute diameter LGS asterism. \textit{Upper}: $C_n^2$ profile evolution over the night. \textit{Middle}: Corresponding tomographic error for different optimisation strategies. Solid and dashed black lines indicate optimal and worst strategies respectively, and are the same for all panels. \textit{Lower}: Optimisation times for each strategy, colours same as the middle panels.}
    \label{fig:night44}
\end{figure*}

\begin{figure*}
    \centering
    \includegraphics{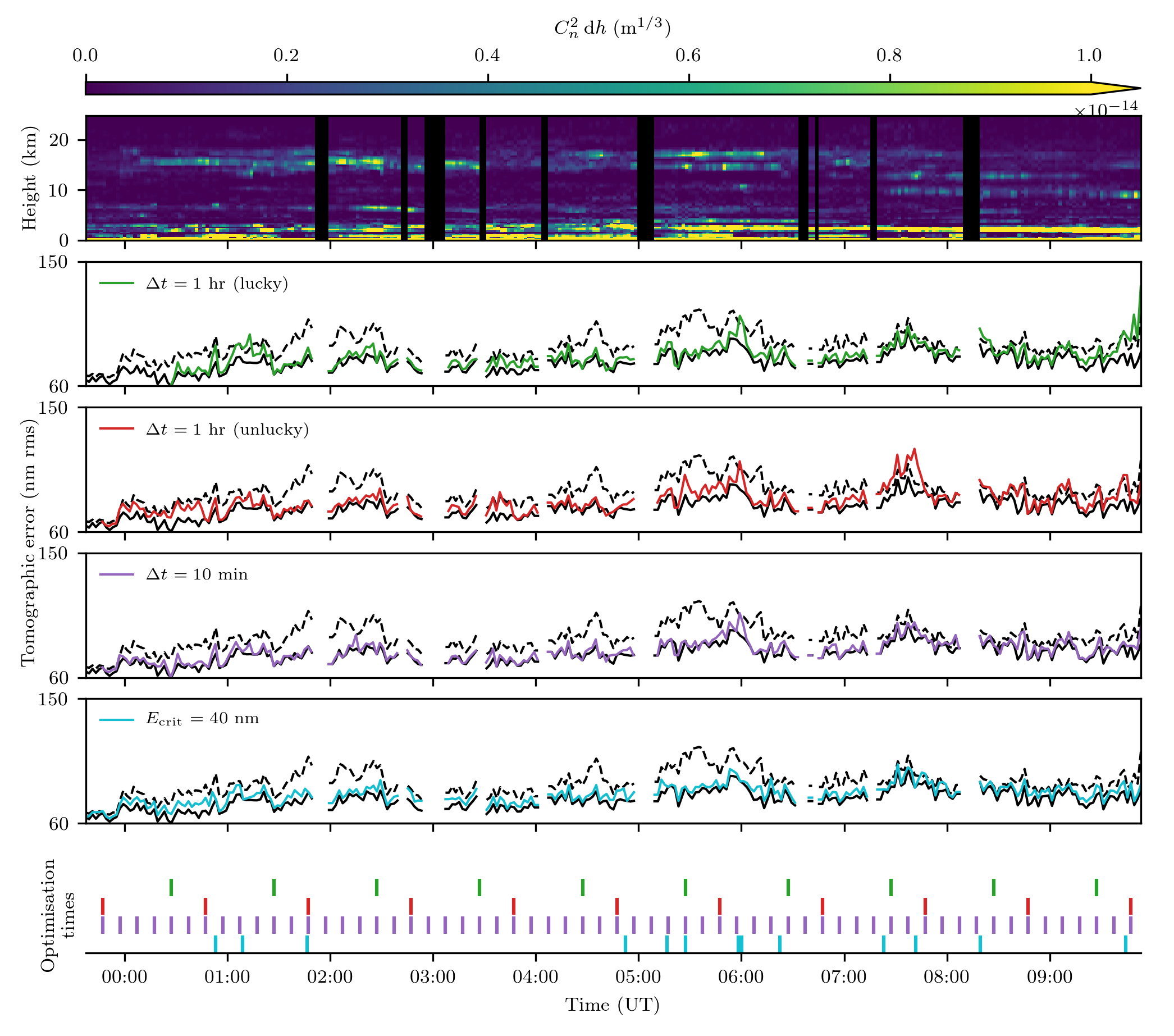}
    \caption{ELT tomographic error over the night of 29th April 2016, using a 1 arminute diameter LGS asterism. Layout as Fig. \ref{fig:night44}.}
    \label{fig:night3}
\end{figure*}

To investigate the differences between these strategies we select two contrasting nights of $C_n^2(h)$ profiles from the dataset. The first, shown in Fig. \ref{fig:night44}, displays a large amount of variability in the profile throughout the night. In the second night in Fig. \ref{fig:night3} the profile is less variable.

For each night, we compare the following optimisation strategies:
\begin{itemize}
    \item \textbf{Optimal strategy}: Reconstructor updated with every new profile. Best possible tomographic error.
    \item \textbf{Worst strategy}: Not updating the reconstructor, using only the ESO 35 layer median profile. Corresponds to worst case performance as per Fig. \ref{fig:R_mean}.
    \item \textbf{{$\boldsymbol{\Delta t}$} = 1 hour (lucky)}: Optimising when $t-t_\mathrm{opt} > 1$ hour, at lucky times where the profile does not change much over the optimisation period.
    \item \textbf{$\boldsymbol{\Delta t}$ = 1 hour (unlucky)}: Optimising when $t-t_\mathrm{opt} > 1$ hour, at unlucky times where the profile changes over the optimisation period.
    \item \textbf{$\boldsymbol{\Delta t}$ = 10 minutes}: Optimising when $t-t_\mathrm{opt} > 10$ minutes.
    \item \textbf{$\boldsymbol{E_\mathrm{crit}}$ = 40 nm}: Optimising when the increase in tomographic error becomes greater than 40 nm rms.
\end{itemize}
Of course in reality it is not possible to choose lucky or unlucky times to optimise the reconstructor. However, by simulating all possible optimisation times and selecting the best and worst cases, we show the potential variability than can arise as a result of changing optimisation times.

For the sake of brevity, we perform this analysis only for the 1 arcminute LGS asterism case. Additionally, we assume that sufficient layers are reconstructed that model error is small in comparison to the temporal error. From Figs. \ref{fig:nlayers} and \ref{fig:nlayers_required} this corresponds to an assumption that $N \geq 10$ for a 1 arcminute asterism. Since we have shown in section \ref{sec:avgtime} that the averaging time $\delta t$ has little effect on the tomographic error, we may neglect it here. We also assume that other factors such as changing zenith angle are negligable, i.e. we are pointing near zenith. 

Starting with Fig. \ref{fig:night44}, we see that there are some spikes in error throughout the night. These occur when the profile undergoes a large change and the system is not reoptimised. The worst error spikes occur unsurprisingly for the $\Delta t=1$ hour case, particularly if we optimise at unlucky times. The worst spike at 04:00 UT is a good example of a lucky vs. unlucky optimisation. In the lucky case we happen to optimise just after the profile changes, meaning that we obtain near optimal error around 70 nm rms as opposed to 150 nm if we are unlucky. Clearly if the profile is this variable there is value in optimising at lucky times. Both $\Delta t = 10$ minutes and $E_\mathrm{crit}=40$ nm maintain almost optimal error over the entire night. However, looking at the optimisation times it is clear especially for the first half of the night that $\Delta t=10$ minutes results in some unnecessary reconstructor optimisations.

Our second chosen night in Fig. \ref{fig:night3} shows smaller errors as would be expected from a calmer atmosphere. We do not see the same spikes in error as in Fig. \ref{fig:night44} and for the most part all optimisation strategies perform fairly well. At times such as this when the atmosphere is less variable optimising on timescales of 1 hour is enough to give good performance. The gain from decreasing this timescale to 10 minutes is small. The optimisation times for the $E_\mathrm{crit}=40$ nm case in particular shows how little the profile is changing. For example, the same reconstructor is used for 3 hours between 01:50 and 04:50 UT without any significant increase in error. 

\begin{figure}
    \centering
    \includegraphics{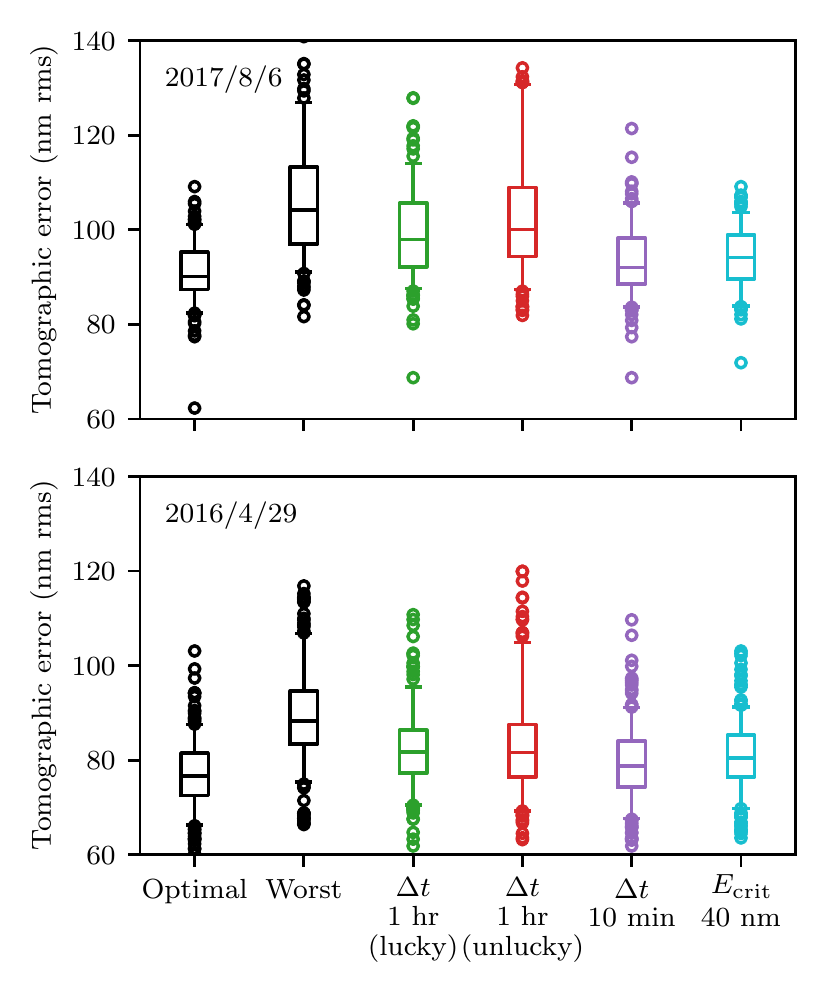}
    \caption{Distribution of increase in tomographic error over the nights of 6th August 2017 (upper panel) and 29th April 2016 (lower panel), for different optimisation strategies. Upper and lower boxplot whiskers represent the 5th and 95th percentiles. Circular markers represent outlier data points above the 95th and below the 5th percentiles.}
    \label{fig:nights_perf}
\end{figure}

Finally, in Fig. \ref{fig:nights_perf} we compare integrated tomographic error across the two nights. Overall error is predictably worse for the first night where the atmosphere is more variable. We see for both nights the closest to optimal tomographic error is obtained by either optimising fast ($\Delta t = 10$ minutes) or by limiting the increase in tomographic error to $E_\mathrm{crit}=40$ nm. However for the more variable night even with 10 minutes between optimisations there are still a small number of outliers pushing up to over 120 nm rms tomographic error, suggesting that despite this fast update rate we can still be unlucky in a small number of cases. For the longer $\Delta t=1$ hour periods, we see the pronounced difference in the upper end of the error distribution for the more variable night indicating the gain from optimising at the right (lucky) times.

Thus it seems that optimising to some maximum error threshold, allowing for the reconstructor optimisation period to dynamically change over the course of a night, gives the greatest flexibility. When the atmosphere is calm, the reconstructor does not need to be optimised for multiple hours, however when an unpredictable change in the profile occurs we can quickly reoptimise and avoid some of the larger increases in error that we see in the spikes of Fig. \ref{fig:night44}. The exact value of $E_\mathrm{crit}$ for a particular system will need to be tuned depending on the tolerances in the error budget and the capabilities of the SRTC.

\section{Conclusions} \label{sec:conclusion}

We have shown using fast AO simulation the effects of sub-optimal tomographic reconstruction on AO performance for an ELT-scale system with a large database of real turbulence profiles from the Stereo-SCIDAR at ESO Paranal.

The number of reconstructed layers, $N$, can have a significant impact on the tomographic error when below a certain threshold. The exact number of layers required will depend on the tolerated level of error for a particular instrument and the LGS asterism diameter. The best tomographic error for a given number of layers is obtained with the optimal grouping compression method. Using this as a baseline, 2 - 6 addditional equivalent layers are required to achieve the same error. If the layers are fixed in altitude, a further 2 - 10 layers are required. Variability of turbulence profiles also plays a role here, as some profiles lend themselves to modelling with few layers whereas others do not. A system wishing to operate to a given error tolerance in the best 95\% of turbulence profiles (with respect to the tomographic error) will need to reconstruct between 6 - 12 additional layers compared to a system operating only in the best 50\% of profiles.

The increase in tomographic error over time was investigated, with the temporal sampling of the Stereo-SCIDAR data allowing us to probe scales as small as several minutes. We find that, although the absolute increase in error is greater for larger asterisms, the shape of this increase with time is similar for all asterisms. After around 20 minutes, the increase in error plateaus and we see smaller increases in tomographic error with time. We therefore conclude that the scale of temporal atmospheric variations as seen by a tomographic AO system is of the order of 20 minutes, and as such one should choose a reconstructor update period of at least $\Delta t < 20$ minutes. The increase in tomographic error after 1 hour, is on average between 30 and 130 nm rms depending on LGS asterism. However, the unpredictable variability of the profile means that large spikes in performance can occur on minute timescales, making it difficult to select a single optimal update rate. 

We also investigated the effect of averaging time $\delta t$. The error arising from increased averaging of the profile is linked to the reconstructor update rate $\Delta t$. For small $\Delta t$, increasing $\delta t$ only makes tomographic error worse: the averaged profile looks less like the real profile. However, if we allow the profile to evolve for longer before reoptimising by increasing $\Delta t$, averaging is less important. There is in fact some optimum value of $\delta t>\delta t_\mathrm{min}$ that slightly improves the tomographic error. This means that a small amount of averaging, usually of the order of $\delta t=10$ minutes, can give the reconstructor slightly more resilience to a changing profile, but only if we are reoptimising on long timescales $\Delta t>20$ minutes. The gain by averaging is very small (of the order of a few percent) compared to the other sources of error investigated here. From a purely atmospheric perspective, i.e. considering only the non-stationary nature of the profile, averaging does not confer any advantage or disadvantage.

Finally, we selected two contrasting nights to compare temporal optimisation strategies. When optimising on long timescales ($\Delta t=1$ hour ) it was found, particularly on the night where the profile is more variable, that it is important to be lucky. That is, to optimise the reconstructor at the right times to avoid large tomographic error spikes. Since this is impossible in reality a very short optimisation period of the order of $\Delta t = 10$ minutes must be used. On the night where the profile is less variable, however, optimising only once per hour gives good results. By optimising only when the increase in tomographic error reaches some threshold $E_\mathrm{crit}$, we are able to obtain good tomographic error at all times by allowing the optimisation period to change dynamically with the atmosphere. With careful selection of $E_\mathrm{crit}$ for a system, the requirements for the SRTC can be relaxed whilst maintaining near optimal tomographic reconstruction.

\section*{Acknowledgements}

This work was supported by the Science and Technology Funding Council (UK) (ST/P000541/1) and (ST/PN002660/1). OJDF acknowledges the support of STFC (ST/N50404X/1).

Horizon 2020: This project has received funding from the European Union's Horizon 2020 research and innovation programme under grant agreement No 730890. This material reflects only the authors views and the Commission is not liable for any use that may be made of the information contained therein.

CC received support from A*MIDEX (project no. ANR-11- IDEX-0001- 02) funded by the ``Investissements d'Avenir" French Government program, managed by the French National Research Agency (ANR).

This work was supported by the Action Sp\'ecifique Haute R\'esolution Angulaire (ASHRA) of CNRS/INSU co-funded by CNES.

This work has been partially funded by the ANR program APPLY - ANR-19-CE31-0011. 

JO acknowledges supoort from the UKRI Future Leaders Fellowship (UK) (MR/S035338/1).

This research made use of Python including NumPy and SciPy \citep{VanderWalt2011}, Matplotlib \citep{Hunter2007} and Astropy, a community-developed core Python package for Astronomy \citep{Robitaille2013}. We also made use of the Python AO utility library AOtools \citep{Townson2019}.

We would like to thank A. Tokovinin for his thorough and insightful review of this work. 



\bibliographystyle{mnras}
\bibliography{library} 








\bsp	
\label{lastpage}
\end{document}